\documentclass[final,preprint,aps,showpacs,tightenlines,superscriptaddress,nofootinbib]{revtex4} 
\usepackage{graphicx}
\usepackage{bm}
\usepackage{dcolumn}

\begin{document}

\bibliographystyle{prsty}

%%%%%%%%%%%%%%%%%%%%%%%%%%%%%%%%%%%%%%%%%%%%%%%%%%%%%%%%%%%%%%%%%%%%%%%%%%%

\preprint{\vbox{\hbox{PNU-NTG-03/2002} \hbox{RUB-TP2-11-02}}}
\title{Strange and singlet form factors of the nucleon: Predictions
for G0, A4, and HAPPEX-II experiments}

\author{A. Silva$^{(1,2)}$
\footnote{E-mail address: ajose@teor.fis.uc.pt},
H.-Ch. Kim$^{(3)}$
\footnote{E-mail address: hchkim@pusan.ac.kr},
and K. Goeke$^{(1)}$
\footnote{E-mail address: Klaus.Goeke@tp2.ruhr-uni-bochum.de}}

\affiliation{(1) Institut f\"ur Theoretische  Physik  II, \\  
Ruhr-Universit\" at Bochum, \\
 D--44780 Bochum, Germany  \\
(2) Departamento de F\'\i sica and Centro de F\'\i sica Computacional,\\
Universidade de Coimbra,\\
P-3000 Coimbra, Portugal\\
(3) Department of Physics and Nuclear Physics \& Radiation Technology
Institute (NuRI), Pusan National University,\\
609-735 Busan, Republic of Korea }
\date{June 2003}

\begin{abstract}
We investigate the strange and flavor-singlet electric and magnetic 
form factors of the nucleon within the framework of the SU(3) chiral 
quark-soliton model.  Isospin symmetry is assumed and the
symmetry-conserving SU(3) quantization is employed, rotational and
strange quark mass corrections being included.  
For the experiments G0, A4, and HAPPEX-II we
predict the quantities $G^{0}_E + \beta G^{0}_M$ and $G^{\rm s}_E + 
\beta G^{\rm s}_M$.  The dependence of the results on the parameters 
of the model and the treatment of the Yukawa asymptotic behavior of the
soliton are investigated.  
\end{abstract}

\maketitle
%\tableofcontents

\section{Introduction}
It is of utmost importance to understand the strangeness content of the
nucleon, since it gives a clue about its internal quark structure.
In particular, the deviation from the valence quark
picture and the polarization of the quark sea must be investigated.
In fact, with this aim a great deal of experimental and theoretical
effort has been put into the study of the strangeness in the nucleon
in various channels: The spin content of the
nucleon~\cite{EMC,SMC1,SMC2,E142,E143},  
the $\pi N$ sigma term $\Sigma_{\pi N}$~\cite{Gasser:1991ce}, 
and the strange vector form factors~\cite{Kaplan:1988ku,Jaffe:1989mj}.  
In particular, the strange vector form factors have been a hot issue recently,
as their first measurement was achieved by the SAMPLE
collaboration~\cite{Mueller:1997mt,Spayde:2000qg,SAMPLE00s} at
MIT/Bates, parity-violating electron scattering being used.
The most recent result by the SAMPLE collaboration~\cite{SAMPLE00s} for the 
strange magnetic form factor finds ($Q^2$ in (GeV/$c$)$^2$) 
\begin{equation} 
        G_M^{\rm s} (Q^2=0.1) = ( + 0.14 \pm 0.29 \, ({\rm stat.}) 
\pm 0.31 \, ({\rm syst.})) \;\; \mbox{n.m.}\, .  
\label{eq:samplenew}
\end{equation}
It is extracted from the knowledge of both the neutral weak magnetic 
form factor $G_M^Z$ measured in parity-violating elastic e-p
scattering and the electromagnetic form factors $G_{M}^{p\gamma}$,  
$G_{M}^{n\gamma}$ by using the relation (assuming isospin invariance)
\begin{equation}
G_M^Z =  \left(1-4\sin^2\theta_W\right)  G_M^{p\gamma} 
- G_M^{n\gamma} - G_M^{s}\, ,       
\label{eq:GMZ}
\end{equation} 
where $\theta_W$ is the Weinberg mixing angle determined  
experimentally~\cite{PDG00}: $\sin^2\theta_W=0.23147$.

The HAPPEX collaboration at TJNAF also announced the measurement of 
the strange vector form factors~\cite{Aniol:2000at}.  The asymmetry
$A_{\rm th}$ is obtained from the parity-violating polarized
electron scattering, from which the singlet form factors are
extracted: 
\begin{equation}
 \frac{(G_E^{0} + 0.392 G_M^{0})}{(G_M^{p\gamma }/\mu_p)} (Q^2=0.477)
        =  1.527\pm 0.048\pm 0.027\pm 0.011.
\label{eq:happex0}
\end{equation}
With the help of the available data for the electromagnetic form
factors via the relation
\begin{equation}
G_{E,M}^{\rm s}  =  
G_{E,M}^0  -  G_{E,M}^{p\gamma}  -  G_{E,M}^{n\gamma},
\label{eq:GEMs}
\end{equation}
the HAPPEX Collaboration arrives at the following result about the
strange form factors: 
\begin{equation}
 (G_E^{\rm s} + 0.392 G_M^{\rm s})(Q^2=0.477)  =  0.025\pm 0.020 \pm 0.014,
\label{eq:happexs}
\end{equation}
where the first error is experimental and the second one is from the 
uncertainties in electromagnetic form factors.  
 
There has been a great deal of theoretical effort in order to predict
the strange vector form factors~\cite{Theory} and each approach emphasizes     
different aspects.  Beck, Holstein, and McKeown reviewed some of
theoretical works in Refs.~\cite{Beck:2001dz,Beck:2001yx}.    

A proper theoretical description of the strange form factors of the
nucleon should be based on QCD.  Since, however, these form factors
basically reflect the excitation of $\rm s\bar{s}$ pairs, it is very
difficult to use lattice gauge techniques because they are still
hampered by technical problems, in particular, with light quarks.
Thus, appropriate models are required, which are based on QCD and
treat the relevant degrees of freedom in a good approximation.  One of
those is the chiral quark-soliton model ($\chi$QSM).  It is an
effective quark theory of the instanton-degrees of freedom of the QCD
vacuum and results in a Lagrangian for valence and sea quarks both
moving in a static self-consistent Goldstone background 
field~\cite{Christov:1996vm,Alkofer:1994ph}.  It has successfully been  
applied to electromagnetic and axial-vector form
factors~\cite{Christov:1996vm} and to forward and
generalized parton distributions~\cite{Diakonov:1996sr,Petrov:1998kf,
Goeke:2001tz} and has lead even to the prediction of the heavily
discussed pentaquark baryon $\Theta^+$~\cite{Diakonov:1997mm}.    

Two of the present authors studied the strange vector form factors within
the framework of this $\chi$QSM some years ago~\cite{Kim:1996vj}.
The formalism used contains a conceptual difficulty because the rotational
corrections break the venerable Gell-Mann-Nishijima
relation~\cite{Watabe:1998vi}.  The reason for this lies probably in
the fact that the large $N_c$ expansion, underlying the stationary
phase approximation of the $\chi$QSM, has not been fully extended to
the SU(3) collective quantization procedure.  Instead, Prasza{\l}owicz
{\em et al.}~\cite{Praszalowicz:1998jm} suggested on practical ground
an approximate formalism, which fulfills the Gell-Mann-Nishijima
relation and in addition has proper limits for large and small
solitonic radii, {\em i.e.} the limit of the Sykrme and the
nonrelativistiv quark model, respectively.  This so-called
symmetry-conserving quantization method~\cite{Praszalowicz:1998jm} is
used in the present paper and the corresponding formulae, correcting
also a technical error of Ref.~\cite{Kim:1996vj}, are given in the
appendix. 

The present authors have recently reinvestigated the strange vector
form factors, following the above-mentioned quantization scheme
suggested by Prasza{\l}owicz {\em et al.}~\cite{Praszalowicz:1998jm}.
We presented some aspects of the SAMPLE, HAPPEX, and A4 experiments
within the framework of the $\chi$QSM~\cite{Silva:2001st}.  Results
have shown a fairly good agreement with experimental data of the
SAMPLE and HAPPEX.  In this work, we want to extend the former
investigation, to document the relevant formulae of the model and to
present the results pertinent to future experiments: G0 experiments
being conducted at TJNAF will measure a linear combination of the strange
vector form factors at seven different values of the momentum transfer
$Q^2$ with two different angles, {\em i.e.} the forward angle
$\theta=10^\circ$ and the backward angle $108^\circ$.  With these
measurements, the strange electric and magnetic form factors can be
separately obtained.  The A4 experiment at MAMI will soon bring out
the new data at $Q^2=0.227\ {\rm GeV}^2$ with
$\theta=35^\circ$~\cite{Maas:2003xp}.  The 
planned HAPPEX II experiment will measure the combination of the
strange vector form factors at $Q^2=0.11\ {\rm GeV}^2$, which is the
same momentum transfer as the SAMPLE experiment, with the forward
angle $\theta=6^\circ$ to extract the separated strange electric and 
magnetic form factors with the SAMPLE data combined.  Thus, in the
present work, we will continue our previous work~\cite{Silva:2001st}
and will concentrate on predicting the above-given future experiments,
in particular, G0 experiment.

\section{Strange and singlet vector form factors}
In this section we very briefly review the formalism of the $\chi$QSM.    
Details of the model~\cite{Diakonov:1988ty} can be found in 
ref.~\cite{Christov:1996vm}.  Employing in the following the
non-standard sign convention  used by Jaffe~\cite{Jaffe:1989mj} for
the strange vector current, the strange and singlet vector form
factors of the baryons are expressed in the quark matrix elements as
follow: 
\begin{equation}
\langle N (p')|J^{{\rm s},(0)}_{\mu }| N(p)\rangle \; =\; \bar{u}_{N}(p')
\left[ \gamma _{\mu }F^{{\rm s},0}_{1}(q^{2})+i\sigma _{\mu \nu }
\frac{q^{\nu }}{2M_{N}}F^{{\rm s},0}_{2}(q^{2})\right] u_{N}(p),
\label{Eq:ff1}
\end{equation}
 where \( q^{2} \) is the square of the four momentum transfer 
\( q^{2}=-Q^{2} \) with \( Q^{2}>0 \). \( M_{N} \) and \( u_{N}(p) \) 
stand for the nucleon mass and its spinor, respectively. The strange quark 
current $J^{\rm s}_{\mu }$ can be expressed in terms of the
flavor-singlet and flavor-octet currents in Euclidean space: 
\begin{equation}
\label{Eq:scur}
J^{\rm s}_{\mu } \;=\; -i\psi^\dagger \gamma _{\mu }\hat{Q}_{s} \psi
\; =\; \frac{1}{N_{c}}J^{(0)}_{\mu }-\frac{1}{\sqrt{3}}J^{(8)}_{\mu },
\end{equation}
where $J^{(0)}_{\mu }$ and $J^{(8)}_{\mu }$ are the flavor-singlet
and flavor-octet currents, respectively: 
\begin{eqnarray}
J^{(0)}_{\mu } & = & -i\psi^\dagger\gamma _{\mu }\psi
\nonumber \\
J^{(8)}_{\mu } & = & -i\psi^\dagger \gamma _{\mu }\lambda _{8}\psi.
\label{Eq:bar}
\end{eqnarray}
Here, $N_{c}=3$ is correctly introduced to make it sure that the
baryon number must be equal to unity.  The baryon and hypercharge
currents are equal to the singlet and octet currents, respectively. 

The strange (singlet) Dirac form factors 
$F^{{\rm s},0}_{1}$ and $F^{{\rm s},0}_{2}$ can be
written in terms of the strange (singlet) Sachs form factors, 
$G^{{\rm s},0}_{E}(Q^{2})$ and $G^{{\rm s},0}_{M}(Q^{2})$: 
\begin{eqnarray}
G^{{\rm s},0}_{E}(Q^{2}) & = & F^{{\rm s},0}_{1}(Q^{2})
-\frac{Q^{2}}{4M^{2}_{N}}F^{{\rm s},0}_{2}(Q^{2})\nonumber \\
G_{M}^{{\rm s},0}(Q^{2}) & = & F^{{\rm s},0}_{1}(Q^{2})
+F^{{\rm s},0}_{2}(Q^{2}).
\end{eqnarray}

Having carried out a lengthy calculation following
strictly Refs.\cite{Kim:1995mr,Praszalowicz:1998jm}, we obtain the
expressions for the strange vector form factors and flavor-singlet
form factors of the nucleon: 
\begin{eqnarray}
G_{E}^{{\rm s},0} ({\bm Q}^2) &=& 
G_{E}^{{\rm s},0,m_s^{0}} ({\bm Q}^2) + 
G_{E}^{{\rm s},0,m_{\rm s}^{1},{\rm op}} ({\bm Q}^2) +
G_{E}^{{\rm s},0,m_{\rm s}^{1},{\rm wf}} ({\bm Q}^2),\nonumber \\
G_{M}^{{\rm s},0}({\bm Q}^2)&=& 
G_{M}^{{\rm s},0,m_s^{0}} ({\bm Q}^2) + 
G_{M}^{{\rm s},0,m_{\rm s}^{1},{\rm op}} ({\bm Q}^2) +
G_{M}^{{\rm s},0,m_{\rm s}^{1},{\rm wf}} ({\bm Q}^2),
\label{Eq:final}
\end{eqnarray}
where $G_{E}^{{\rm s},m_{\rm s}^{0}} ({\bm Q}^2) (G_{M}^{0,
m_{\rm s}^{0}} ({\bm Q}^2))$ 
stands for the SU(3) symmetric part of the strange (flavor-singlet) 
electric and magnetic form factors, whereas the symmetry breaking parts 
$G_{E}^{{\rm s},m_{\rm s}^{1},{\rm op}} ({\bm Q}^2)
(G_{M}^{0,m_{\rm s}^{1},{\rm op}} ({\bm Q}^2))$ and 
$G_{E}^{{\rm s},m_{\rm s}^{1},{\rm wf}} ({\bm Q}^2)
(G_{M}^{0,m_{\rm s}^{1},{\rm wf}} ({\bm Q}^2))$ correspond to 
the symmetry breaking in the operator and in the baryon wave functions,
respectively.  The explicit expressions for
the strange vector form factors in Eq.(\ref{Eq:final}) are given
below.  They differ from those of Ref.~\cite{Kim:1996vj} by some
numerical constants and by discarding some redundant terms.
\begin{eqnarray}
G_{E}^{{\rm s},m_{\rm s}^{0}} ({\bm Q}^2) &=& 
\frac1{10}\left(7 {\cal B}({\bm Q}^2)
- \frac{{\cal I}_1 ({\bm Q}^2)}{I_1}-6\frac{{\cal I}_2 ({\bm
Q}^2)}{I_2} \right) \label{es0B8} \\
&&  \nonumber \\
G_{E}^{{\rm s},m_{\rm s}^{1},{\rm op}} ({\bm Q}^2) &=&\frac 1{15}\left(
(m_0 - \bar{m})13
+m_8\frac{5}{\sqrt{3}}\right) {\cal C}({\bm Q}^2) \nonumber \\
&+&m_8 \frac{12}{15\sqrt{3}}\left(I_1 {\cal K}_1 ({\bm Q}^2 ) - 
K_1{\cal I}_1({\bm Q}^2)\right) \nonumber \\
&+& m_8\frac{12}{5\sqrt{3}} \left(I_2 {\cal K}_2 ({\bm Q}^2 ) - 
K_2{\cal I}_2({\bm Q}^2)\right) \label{es1opB8} \\
&&  \nonumber \\
G_{E}^{{\rm s},m_s^{1},{\rm wf}} ({\bm Q}^2) &=&
-m_8 \left(c_{\overline{10}} + \frac{6\sqrt{3}}{5I_1} \, 
c_{27}\right)\,{\cal B}({\bm Q}^2 )  
\nonumber \\  
&-& m_8\frac{1}{5I_1} \left(5 \,c_{\overline{10}} - 6
  \,c_{27}\right)\, {\cal I}_1 ({\bm Q}^2)
- m_8 \frac {24}{5\sqrt{3} I_2}c_{27} {\cal I}_2 ({\bm Q}^2), \\
G_{M}^{{\rm s},m_{\rm s}^{0}} ({\bm Q}^2) &=&\frac{M_N}{3|{\bm
    Q}|}\left\{-\frac 8{30} \left({\cal Q}_0 ({\bm Q}^2 ) + 
\frac{1}{I_1} {\cal Q}_1 ({\bm Q}^2 )  + \frac{1}{6I_2} {\cal X}_2({\bm Q}^2) 
\right)\,S_3 \right.\cr
&-&\left.\frac{1}{15I_1}{\cal X}_1 ({\bm Q}^2)\,S_3 \right\}  \\
G_{M}^{{\rm s},m_{\rm s}^{1},{\rm op}} ({\bm Q}^2) &=& 
\frac{M_N}{3|{\bm Q}|}\left\{-m_8\frac 4{135} \left(6{\cal M}_2 ({\bm Q}^2) - 2
    \frac{K_2}{I_2} {\cal X}_2 ({\bm Q}^2) \right)\,S_3\right. \cr
&-& m_8\frac1{9}\left({\cal M}_0 ({\bm Q}^2) 
+ {\cal M}_1 ({\bm Q}^2)
-\frac13\frac{K_1}{I_1}{\cal X}_1 ({\bm Q}^2)\right)\,S_3 \cr
&-& m_8\frac 1{15} \left({\cal M}_0 ({\bm Q}^2) - {\cal M}_1 ({\bm Q}^2)
+\frac13\frac{K_1}{I_1}{\cal X}_1 ({\bm Q}^2)\right)\,S_3\cr
&+& \left.(m_0 - \bar{m})\frac 8{15} {\cal M}_0 ({\bm Q}^2)\,
S_3\right\},  \label{ms1opB8} \\
G_{M}^{{\rm s},m_s^{1},{\rm wf}} ({\bm Q}^2) &=&\frac{M_N}{3|{\bm
    Q}|}\left\{-m_8\frac 8{45}\,c_{27}\,
\left({\cal Q}_0 ({\bm Q}^2 ) + 
\frac{1}{I_1} {\cal Q}_1 ({\bm Q}^2 )   \right.\right.\cr
&+& \left.\left. \frac{2}{I_2} {\cal X}_2
({\bm Q}^2)-\frac{3}{2I_1} {\cal X}_1 ({\bm Q}^2) \right)\right\} \,S_3.
\label{Eq:semform}
\end{eqnarray}
The flavor-singlet vector form factor are written as:
\begin{eqnarray}
G_{E}^{0,m_{\rm s}^{0}} ({\bm Q}^2) &=& 
{\cal B}({\bm Q}^2), \label{e0B8} \\
&&  \nonumber \\
G_{E}^{{0},m_{\rm s}^{1},{\rm op}} ({\bm Q}^2) &=& \left(
2 (m_0 - \bar{m})
+m_8\frac{3}{10\sqrt{3}}\right) {\cal C}({\bm Q}^2), \label{e1opB8} \\ 
G_{E}^{0,m_s^{1},{\rm wf}} ({\bm Q}^2) &=& 0, \\
G_{M}^{{0},m_{\rm s}^{0}} ({\bm Q}^2) &=&\frac{M_N}{|{\bm
    Q}|}\frac{{\cal X}_1 ({\bm Q}^2)}{I_1}\,S_3  \\
G_{M}^{{0},m_{\rm s}^{1},{\rm op}} ({\bm Q}^2) &=& 
\frac{M_N}{|{\bm Q}|}\frac{m_8\sqrt{3}}{15}
\left(6{\cal M}_1 ({\bm Q}^2) - 2
    \frac{K_1}{I_1} {\cal X}_1 ({\bm Q}^2) \right)\,S_3\\
G_{M}^{{0},m_s^{1},{\rm wf}} ({\bm Q}^2) &=& 0,
\label{Eq:singform}
\end{eqnarray}
where the coefficients like $c_{\overline{10}}$ are
known from the SU(3) algebra, the $I_1$ etc. are 
moments of ineria whose expressions can be found in Ref.
\cite{Blotzetal}.  The other quantities like ${\cal I}_1 ({\bm Q}^2)$
are explicitely given in appendix A.  

\section{Results and discussion}
We present now the results obtained from the $\chi$QSM.  A detailed
description on numerical methods is presented in 
Refs.~\cite{Kim:1995mr,Christov:1996vm}.  The parameters of the model 
are the constituent quark mass $M$, the current quark mass $m_{\rm u}$, the
cut-off $\Lambda$ of the proper-time regularization, and the strange quark mass
$m_{\rm s}$. These parameters are not free but have to be adjusted to
independent observables in a very clear way (in fact this was done
many years ago): The $\Lambda$ and the
$m_{\rm u}$ are  adjusted for a given $M$ in the mesonic sector.  The
physical pion mass $m_\pi = 139$ MeV and the pion decay constant
$f_\pi = 93$ MeV are reproduced by these parameters.  The strange
quark mass is chosen  to be $m_{\rm s} = 180$  MeV throughout the
present work.  The remaining parameter $M$ is varied from $400$ MeV to  
$450$ MeV.  The value of $420\mbox{ MeV}$, which for many years is
known to produce  the best fit to many baryonic
observables~\cite{Christov:1996vm}, is selected for our final result  
in the baryonic sector.  The magnetic moments of the proton and
neutron in the symmetry-conserving quantization are: $\mu_{\rm p} =
1.77\,\mu_N$ and $\mu_{\rm n} = -1.20\,\mu_N$, respectively.  Compared to
the experimental data, they are underestimated by $35\,\%$.  

We always assume isospin symmetry.  Actually
with this formalism we obtained the results (within the admittedly
large experimental errors) in fairly good agreement with the data  
of SAMPLE and HAPPEX.

The formalism of the $\chi$QSM has been applied frequently to SU(3)
baryons.  In the present case, where explicitly a strange quantity is
considered, one meets a problem concerning the asymptotic behavior of
the kaon field.  While in SU(2) the soliton incorporates the
asymptotic pion behavior $\exp(-\mu r)/r$ with $\mu=m_\pi$ in a
natural way, the construction of the SU(3) hedgehog by Witten's
embedding causes all other Goldstone bosons to share the same
asymptotic behavior.  This, however, contradicts the common belief
that the asymptotic form of the kaon field is given by 
$\exp(-\mu r)/r$ with $\mu=m_K$.  Therefore, in our procedure, there
is some sort of systematic error which we have to estimate.  We do
this described in Ref.~\cite{Silva:2001st}, performing two separate
calculations, first by choosing the parameter $\bar{m}$ in the
one-bady Dirac Hamiltonian with the background meson fields such that
the SU(3) calculation yields a pionic tail for all the solitonic profiles
($\mu=m_\pi$), and second, by selecting $\bar{m}$ in order to
get a kaonic tail ($\mu=m_K$).  In both cases we compensate by
modifying the perturbative collective treatment of $m_{\rm s}$ (or
$m_8$) by subtracting the corresponding term.  Altogether we get for
each purely strange contribution to an observable two values the
differences of which gives a measure for the systematic error of our
solitonic calculation (for the up and down part of an observable we
always use the Yukawa mass $\mu=m_\pi$).  To get a feeling for the
dependence of our results on the other parameters of the model, we also
present results for various constituent quark mass $M$ (always
yielding a correct $f_\pi = 93\,{\rm MeV}$ and $m_\pi = 139\,{\rm
  MeV}$, of course).       

The magnetic strange form factor $G_{\rm M}^{\rm s}(Q^2)$ for three
different values of $M$ is shown in Fig. 1 with the pionic
asymptotics.  This is a typical case such that altogether the effect
of the variation of $M$ is not very important and smaller than the
effect of having different Yukawa tails.  This quantitative feature is
true for the electric strange form factors as well. Thus, we consider
only the results of our model with $M=420$ MeV, since it has been
shown in several calculations~\cite{Christov:1996vm} that many other
baryonic observables are reproduced well.

The figures for the form factors of the present calculations have been
published already in Ref.~\cite{Silva:2001st}.  For completeness, we
present here the strange Dirac and Pauli form factors in Figs. 2, 3
with kaon and pion tails, respectively.  They will be suitable 
for a direct comparison with the results of the A4 experiment which
will soon come out.       

In Table I we display the prediction of the $\chi$QSM for the G0
experiment.  Presented is the combination of the strange vector
form factors $G_{E}^{{\rm s}} + \beta G_{M}^{{\rm
s}}$.  Here, $\beta$ is defined as 
\begin{equation}
\beta (Q^2,\theta) \;=\; \frac{\tau G_{M}^{p\gamma}}{\epsilon
  G_{E}^{p\gamma}},
\end{equation}
where $\tau = Q^2/(4M_N^{2})$ and $\epsilon = 
[1+2(1+\tau)\tan^2(\theta/2)]^{-1}$ and the $G_M^{p\gamma}$ and
$G_{E}^{p\gamma}$ are taken from the experiment.  For smaller $Q^2$
values, $G_{E}^{{\rm s}} + \beta G_{M}^{{\rm s}}$ is rather sensitive
to which tail we use.  For example, the result at $Q^2=0.16\ {\rm
  GeV}^2$ shows $15$ to $50$ \% difference at $\theta=10^\circ$ and
$\theta=108^\circ$, respectively, between kaon and pion
tails, whereas at $Q^2=0.951\ {\rm GeV}^2$ we find $10$ to $15\, \%$
difference.  Thus, smaller $Q^2$ show more sensitivity to the tail.  The
difference between the results from the kaon and pion tails is 
comparatively smaller at backward angles. In any case this difference
indicates the size of the systematic error of our model.  

In Table II we list the predictions of the singlet form factor
$G_{E}^{0} + \beta G_{M}^0$ for the G0 experiment and in
in Table III and IV we present, respectively, the predictions of the
strange form factor 
$G_{E}^{{\rm s}} + \beta G_{M}^{{\rm s}}$ and singlet form factor
$G_{E}^{0} + \beta G_{M}^{0}$ at $Q^2=0.227{\rm 
GeV}^2$ for the A4 experiment.  In Table V we list 
the prediction of the strange form factor for the HAPPEX II
experiment, whereas in table VI we predict the corresponding singlet
form factor. Again the difference between the 
numbers obtained using a pion and also a kaon tail indicates the
size of the systematic error of our model.

\section{Summary and outlook}
In the present work, we have investigated the strange vector form
factors $G_{E}^{{\rm s}}$ and $G_{M}^{{\rm s}}$ and flavor singlet 
vector form factors within the framework of the SU(3) chiral quark-soliton
model, incorporating the symmetry-conserving quantization.  The
rotational $1/N_c$ and strange quark mass $m_{\rm s}$ corrections
were taken into account.  In order to get a feeling for the systematic
error of our approach in calculating such a sensitive quantity as a
strange form factor, we also have considered two different
asymptotic behaviors of the soliton in such a way that the tails of  
the soliton fall off according to the Yukawa mass of the pion and of
the kaon.  

We first have examined in detail the dependence of the strange form
factors on the constituent quark mass $M$ which is the only free
parameter we deal with.  The dependence on the $M$ turned out rather
mild in general and we chose $M=420 $ MeV for which many other
properties of the nucleon and the hyperons are reproduced. We also have
predicted the combination of the strange form factors, {\em i.e.} 
$G_{E}^{{\rm s}} + \beta G_{M}^{{\rm s}}$ and  $G_{E}^{0} + \beta G_{M}^0$
corresponding to kinematics of three different experiments, that is, the G0,
A4, and HAPPEX II experiments.  

For the presently used chiral quark-soliton model the derivation of
a strange contribution to the electromagnetic form factors of the
nucleon is a rather natural thing, since the theory can be
considered as a many body approach with a polarized Dirac sea.  In
fact, as one finds in other observables of the nucleon, about $5\%-10\%$ 
contribution comes from the strange $\rm s\bar{s}$ excitation of the
quark sea~\cite{Christov:1996vm}.  If one compares the present
approach with others in the literature, one finds a difference insofar
that most of the theories yield a negative strange magnetic moment,
whereas the present one produces a slightly positive
one~\cite{Silva:2001st}.  The reason might lie in the fact that the
present approach is the only one with quarks in a self-consistent
static meson field, with a proper treatment of the symmetries in SU(3)
including rotational corrections.  In particular, the meson field is
closely related to the instanton liquid of the QCD vacuum.  So far the
approach has been successful in SU(2) and here we have a sensitive
test in SU(3).  It is planned also to calculate the asymmetries of
parity-violating electron scattering directly.  For this we need
axial-vector form factors calculated in SU(3) which is presently under
way.   
\vspace{0.5cm}

{\em Note added:} While the present paper was in the refereeing
process, the A4 Collaboration announced their
results~\cite{Maas:2004ta} for $Q^2=0.230({\rm GeV}/)^2$ and
$\theta=35^\circ$.  Apparently our predictions agree within the error
bars with their results on $G_E^{\rm s} + 0.225 G_M^{\rm s}=0.039\pm
0.034$ for $F_1^{\rm s}+0.130F_2^{\rm s}=0.032\pm 0.028$. 

\newpage
\section*{Acknowledgment}
The authors are grateful to F. Maas for useful discussions.
HCK thanks D. von Harrach and M. Pitt for valuable discussions and
comments.  AS acknowledges partial financial support from Praxis
XXI/BD/15681/98.  The work of HCK is supported by the Korea
Research Foundation Grant (KRF--2003--041--C20067).  The work has also
been supported by the BMBF, the DFG, the COSY--Project (J\" ulich) and 
POCTI (MCT-Portugal). 

\begin{appendix}
\section{Densities}
In this appendix, we provide the densities for the strange vector  
and flavor-singlet form factors given in Eqs.(\ref{Eq:semform},
\ref{Eq:singform}).  We list only those, which are different from the
ones in the appendix of ref.~\cite{Kim:1996vj}.  The sums run freely
over all single-quark levels including the valence one, except the sum
over $m_0$, which is restricted to negative-energy orbits:
\begin{eqnarray} {\cal B}({\bm Q}^2) &=& \int d^3x j_0 ( Q r)
\left[\Psi_{\rm val}^\dagger
({\bm x}) \Psi_{\rm val}({\bm x}) - \frac12 \sum_n {\rm sgn}(E_n)
\Psi_n^{\dagger} ({\bm x}) \Psi_n ({\bm x})\right],\nonumber \\  
{\cal C} ({\bm Q}^2) &=& N_c \int d^3x j_0 ( Q r)
\int d^3 y \left[
\sum_{n\neq {\rm val}} \frac{\Psi_{n}^\dagger
({\bm x}) \Psi_{\rm val}({\bm x}) \Psi_n^{\dagger}({\bm y})\beta 
\Psi_{\rm val} ({\bm y})}{E_n - E_{\rm val}} \right. \nonumber \\
& & \hspace{3cm}\;-\;\left. \frac12 \sum_{n,m} 
\Psi_n^{\dagger} ({\bm x}) \Psi_m ({\bm x})
\Psi_m^{\dagger}({\bm y})\beta \Psi_n ({\bm y})
{\cal R}_{\cal M} (E_n, E_m)\right], \nonumber \\
{\cal I}_1({\bm Q}^2) & = & \frac{N_c}{6} 
\int d^3 x j_0 ( Q r) \;\int d^3 y
\left [\sum_{n\neq {\rm val}}\frac{\Psi^{\dagger}_{n} ({\bm x}) {\bm \tau}
\Psi_{\rm val} ({\bm x}) \cdot
\Psi^{\dagger}_{\rm val} ({\bm y}) {\bm \tau} \Psi_{n} ({\bm y})}
{E_n - E_{\rm val}} \right .
\nonumber \\  & & \hspace{3cm} \;+\; \left .
\frac{1}{2}\sum_{n,m}
\Psi^{\dagger}_{n} ({\bm x}) {\bm \tau} \Psi_{m} ({\bm x}) \cdot
\Psi^{\dagger}_{m} ({\bm y}) {\bm \tau} \Psi_{n} ({\bm y})
{\cal R}_{\cal I} (E_n, E_m) \right ],
\nonumber \\
{\cal I}_2({\bm Q}^2) & = &\frac{N_c}{4} 
\int d^3 x j_0 ( Q r) \;\int d^3 y
\left [\sum_{m^{0}}\frac{\Psi^{\dagger}_{m^{0}} ({\bm x}) \Psi_{\rm val} 
({\bm x}) \Psi^{\dagger}_{\rm val} ({\bm y}) \Psi_{m{^0}} ({\bm y})}
{E_{m^{0}} - E_{\rm val}} \right .
\nonumber \\  & & \hspace{3cm} \;+\;\left . \frac{1}{2}\sum_{n,m_0}
\Psi^{\dagger}_{n} ({\bm x}) \Psi_{m^{0}} ({\bm x})
\Psi^{\dagger}_{m^{0}} ({\bm y}) \Psi_{n} ({\bm y})
{\cal R}_{\cal I} (E_n, E_{m^{0}}) \right ],
\nonumber \\
{\cal K}_1({\bm Q}^2) & = & \frac{N_c}{6} 
\int d^3 x j_0 ( Q r) \; \int d^3 y
\left [\sum_{n}\frac{\Psi^{\dagger}_{n} ({\bm x})
{\bm \tau} \Psi_{\rm val} ({\bm x}) \cdot
\Psi^{\dagger}_{\rm val} ({\bm y}) \beta {\bm \tau} \Psi_{n} ({\bm y})}
{E_n - E_{\rm val}} \right .
\nonumber \\  & & \hspace{3cm} \;+\; \left . \frac{1}{2}\sum_{n,m}
\Psi^{\dagger}_{n} ({\bm x}) {\bm\tau} \Psi_{m} ({\bm x}) \cdot
\Psi^{\dagger}_{m} ({\bm y}) \beta {\bm \tau} \Psi_{n} ({\bm y})
{\cal R}_{\cal M} (E_n, E_m)
\right ],
\nonumber \\
{\cal K}_2({\bm Q}^2) & = & \frac{N_c}{4} 
\int d^3 x j_0 ( Q r)\; \int d^3 y
\left [\sum_{m^{0}}\frac{\Psi^{\dagger}_{m^{0}} ({\bm x}) \Psi_{\rm val} 
({\bm x}) \Psi^{\dagger}_{\rm val} ({\bm y}) \beta \Psi_{m{^0}} ({\bm y})}
{E_{m^{0}} - E_{\rm val}} \right .
\nonumber \\  & & \hspace{3cm} \;+\;\left . \frac{1}{2}\sum_{n,m_0}
\Psi^{\dagger}_{n} ({\bm x}) \Psi_{m^{0}} ({\bm x})
\Psi^{\dagger}_{m^{0}} ({\bm y}) \beta \Psi_{n} ({\bm y})
{\cal R}_{\cal M} (E_n, E_{m^{0}}) \right ],\\
{\cal Q}_0({\bm Q}^2)  & = &
\frac{N_c}{3}\int d^3 x \frac{j_1 ( Q r)}{r}\,
\left[ \Psi ^{\dagger}_{\rm val}({\bm x}) \gamma_{5}
\{{\bm r} \times {\bm\sigma} \} \cdot {\bm \tau}
\Psi_{\rm val} ({\bm x}) \right. \nonumber \\ & &
\left . \hspace{1cm} \;-\;
\frac{1}{2}  \sum_n {\rm sgn} (E_n)
\Psi ^{\dagger}_{n}({\bm x}) \gamma_{5}
\{{\bm r} \times {\bm\sigma} \} \cdot {\bm\tau}
\Psi_{n}({\bm x}) {\cal R}(E_n)\right ],
\nonumber \\
{\cal Q}_1({\bm Q}^2) & = &  \frac{iN_c}{2}
\int d^3 x \frac{j_1 ( Q r)}{r}\,
\int d^3 y \nonumber \\  & \times &
\left[\sum_{n}{\rm sgn} (E_n)
\frac{\Psi^{\dagger}_{n} ({\bm x}) \gamma_{5}
\{{\bm r} \times {\bm \sigma} \} \times {\bm\tau}
\Psi_{\rm val} ({\bm x}) \cdot
\Psi^{\dagger}_{\rm val} ({\bm y}) {\bm \tau} \Psi_{n} ({\bm y})}
{E_n - E_{\rm val}} \right .
\nonumber \\  & & %\hspace{1cm}
\;+\; \left . \frac{1}{2} \sum_{n,m}
\Psi^{\dagger}_{n} ({\bm x})\gamma_{5} \{{\bm r} \times {\bm\sigma} \}
\times {\bm\tau}  \Psi_{m} ({\bm x}) \cdot
\Psi^{\dagger}_{m} ({\bm y}) {\bm\tau} \Psi_{n} ({\bm y})
{\cal R}_{\cal Q} (E_n, E_m) \right ],
\nonumber \\
{\cal X}_1({\bm Q}^2)   & = & N_c
 \int d^3 x \frac{j_1 ( Q r)}{r} \,
\int d^3 y \left[\sum_{n}
\frac{\Psi^{\dagger}_{n} ({\bm x})\gamma_{5}
\{{\bm r} \times {\bm\sigma} \}
\Psi_{\rm val} ({\bm x}) \cdot
\Psi^{\dagger}_{\rm val} ({\bm y}) {\bm\tau} \Psi_{n} ({\bm y})}
{E_n - E_{\rm val}} \right .
\nonumber \\  &+ & \left. %\hspace{1cm} \;+\; \left .
\frac{1}{2} \sum_{n,m}
\Psi^{\dagger}_{n} ({\bm x})\gamma_{5} \{{\bm r} \times {\bm\sigma} \}
\Psi_{m} ({\bm x}) \cdot
\Psi^{\dagger}_{m} ({\bm y}) {\bm\tau} \Psi_{n} ({\bm y})
{\cal R}_{\cal M} (E_n, E_m) \right ],
\nonumber \\
{\cal X}_2({\bm Q}^2)   & = & N_c
 \int d^3 x \frac{j_1 ( Q r)}{r}\,
\int d^3 y \left[\sum_{m^0}
\frac{\Psi^{\dagger}_{m^0} ({\bm x})\gamma_{5}
\{{\bm r} \times {\bm\sigma} \} \cdot {\bm\tau}
\Psi_{\rm val} ({\bm x})
\Psi^{\dagger}_{\rm val} ({\bm y}) \Psi_{m^0} ({\bm y})}
{E_{m^0} - E_{\rm val}} \right .
\nonumber \\  & & \hspace{1cm}
\;+\; \left . \sum_{n,m^0}
\Psi^{\dagger}_{n} ({\bm x})\gamma_{5} \{{\bm r} \times {\bm\sigma} \}
\cdot {\bm\tau} \Psi_{m^0} ({\bm x})
\Psi^{\dagger}_{m^0} ({\bm y}) \Psi_{n} ({\bm y})
{\cal R}_{\cal M} (E_n, E_{m^0}) \right ],
\nonumber \\
{\cal M}_0 ({\bm Q}^2) & = &  N_c
\int d^3 x \frac{j_1 ( Q r)}{r}\,
\int d^3 y \left[ \sum_{n}
\frac{\Psi^{\dagger}_{n} ({\bm x}) \gamma_{5}
\{{\bm r} \times {\bm\sigma} \} \cdot {\bm\tau}
\Psi_{\rm val} ({\bm x})
\Psi^{\dagger}_{\rm val} ({\bm y}) \beta \Psi_{n} ({\bm y})}
{E_{n} - E_{\rm val}} \right .
\nonumber \\  & & \hspace{1cm}
\;+\; \left . \frac{1}{2} \sum_{n,m}
\Psi^{\dagger}_{n} ({\bm x})\gamma_{5} \{{\bm r} \times {\bm\sigma} \}
\cdot{\bm\tau}  \Psi_{m} ({\bm x})
\Psi^{\dagger}_{m} ({\bm y})\beta \Psi_{n} ({\bm y})
{\cal R}_{\beta} (E_n, E_m) \right ],
\nonumber \\
{\cal M}_1({\bm Q}^2)  & = & \frac{N_c}{3}
\int d^3 x \frac{j_1 ( Q r)}{r}\,
\int d^3 y \nonumber \\  & \times &
\left[\sum_{n} 
\frac{\Psi^{\dagger}_{n} ({\bm x})\gamma_{5}
\{{\bm r} \times {\bm\sigma} \}
\Psi_{\rm val} ({\bm x}) \cdot
\Psi^{\dagger}_{\rm val} ({\bm y}) \beta {\bm\tau} \Psi_{n} ({\bm y})}
{E_n - E_{\rm val}} \right .
\nonumber \\  &+ & %\hspace{1cm}
\left . \frac{1}{2} \sum_{n,m}
\Psi^{\dagger}_{n} ({\bm x})\gamma_{5} \{{\bm r} \times {\bm\sigma} \}
\Psi_{m} ({\bm x}) \cdot
\Psi^{\dagger}_{m} ({\bm y}) \beta {\bm\tau} \Psi_{n} ({\bm y})
{\cal R}_{\beta} (E_n, E_m) \right ],
\nonumber \\
{\cal M}_2({\bm Q}^2) & = &  \frac{N_c}{3} 
\int d^3 x \frac{j_1 ( Q r)}{r}\,
\int d^3 y \nonumber \\  & \times &
\left[\sum_{m^0}
\frac{\Psi^{\dagger}_{m^0} ({\bm x}) \gamma_{5}
\{{\bm r} \times {\bm\sigma} \} \cdot {\bm\tau}
\Psi_{\rm val} ({\bm x})
\Psi^{\dagger}_{\rm val} ({\bm y})\beta \Psi_{m^0} ({\bm y})}
{E_{m^0} - E_{\rm val}} \right .
\nonumber \\  &+ & %\hspace{1cm}
\left . \sum_{n,m^0}
\Psi^{\dagger}_{n} ({\bm x})\gamma_{5} \{{\bm r} \times {\bm\sigma} \}
\cdot {\bm\tau}  \Psi_{m^0} ({\bm x})
\Psi^{\dagger}_{m^0} ({\bm y}) \beta \Psi_{n} ({\bm y})
{\cal R}_{\beta} (E_n, E_{m^0}) \right ]  .
\label{Eq:mdens}
\end{eqnarray}
The regularization functions in Eq.(\ref{Eq:mdens}) are as follows:
\begin{eqnarray}
{\cal R}_{I} (E_n, E_m) & = & - \frac{1}{2\sqrt{\pi}}
\int^{\infty}_{0} \frac{du}{\sqrt{u}} \phi (u;\Lambda_i)
\left [ \frac{E_n e^{-u E^{2}_{n}} +  E_m e^{-u E^{2}_{m}}}
{E_n + E_m} \;+\; \frac{e^{-u E^{2}_{n}} - e^{-u E^{2}_{m}}}
{u(E^{2}_{n} - E^{2}_{m})} \right ],
\nonumber \\
{\cal R}_{\cal M} (E_n, E_m) & = &
\frac{1}{2}  \frac{ {\rm sgn} (E_n)
- {\rm sgn} (E_m)}{E_n - E_m},\nonumber \\
{\cal R} (E_n) & = & \int \frac{du}{\sqrt{\pi u}}
\phi (u;\Lambda_i) |E_n| e^{-uE^{2}_{n}},
\nonumber \\
{\cal R}_{\cal N} (E_n, E_m) & = &
\frac{1}{2}  \frac{ {\rm sgn} (E_n)
- {\rm sgn} (E_m)}{|E_n| + |E_m|},\nonumber \\
{\cal R}_{\cal Q} (E_n, E_m) & = & \frac{1}{2\pi} 
\int^{1}_{0} d\alpha \frac{\alpha (E_n + E_m) - E_m}
{\sqrt{\alpha ( 1 - \alpha)}}
\frac{\exp{\left (-[\alpha E^{2}_n + (1-\alpha)E^{2}_m]/
\Lambda^{2}_i  \right)}}{\alpha E^{2}_n + (1-\alpha)E^{2}_m},
\nonumber \\
{\cal R}_{\beta}  (E_n, E_m) & = &
\frac{1}{2\sqrt{\pi}} \int^{\infty}_{0}
\frac{du}{\sqrt{u}} \phi (u;\Lambda_i)
\left[ \frac{E_n e^{-uE^{2}_{n}} - E_m e^{-uE^{2}_{m}}}
{E_n - E_m}\right],
\label{Eq:regulm}
\end{eqnarray}
where the cutoff function $\phi(u;\Lambda_i)=\sum_i c_i \theta
\left(u - \frac{1}{\Lambda^{2}_{i}} \right)$ is
fixed by reproducing the pion decay constant and other mesonic properties
\cite{Christov:1996vm}.
\end{appendix}

\newpage
%%%%%%%%%%%%%%%%%%%%%%%%%%%%%%%%%%%%%%%%%%%%%%
\centerline{\large \bf FIGURES}
\vspace{1.2cm}

\includegraphics[height=12cm]{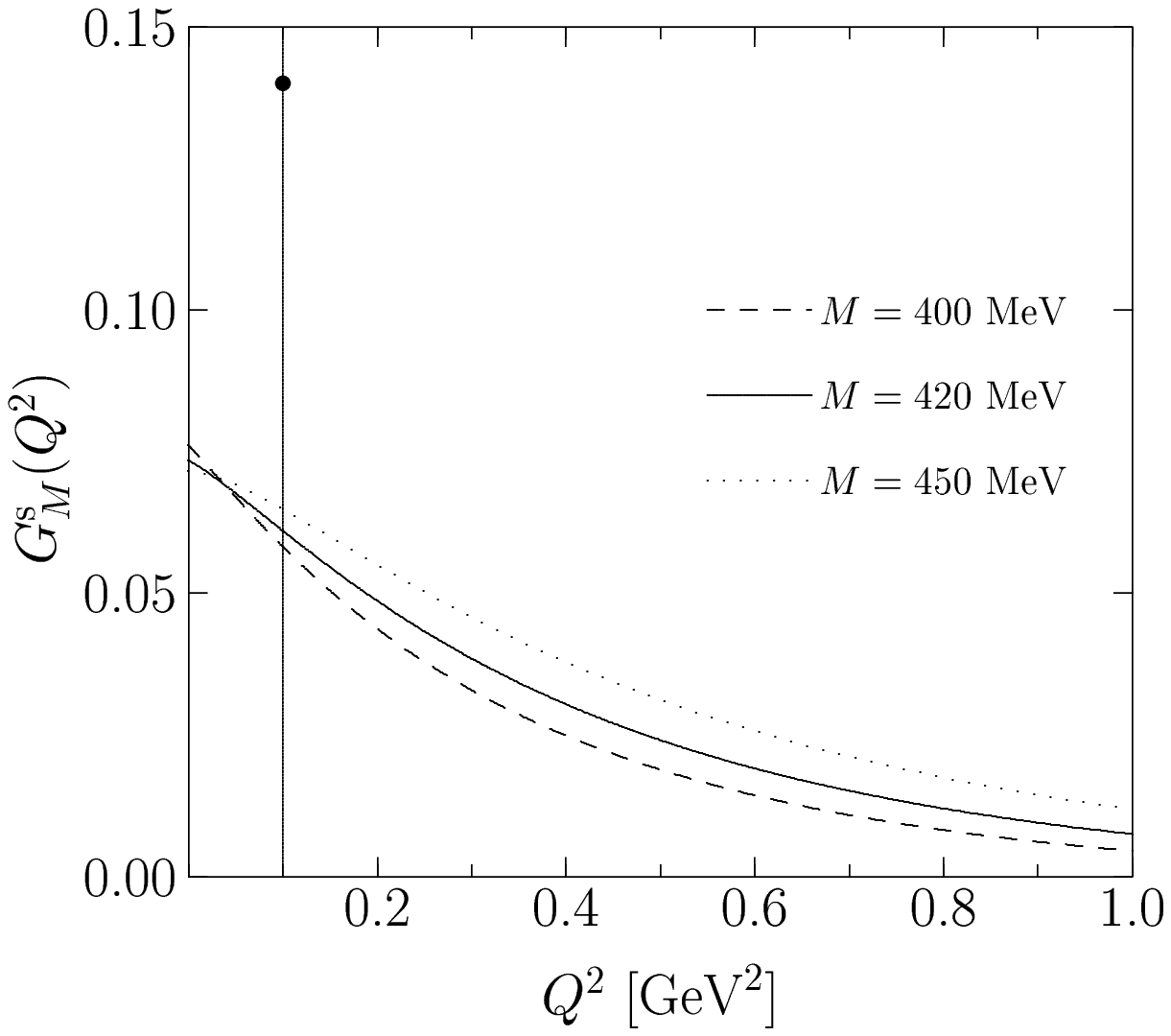}
\vspace{0.8cm}

\noindent {\bf FIG.1}: The dependence of the strange magnetic form
factor as a function of $Q^2$ on the constituent quark mass with the
pion asymptotic tail ($\mu=140$ MeV).
The solid curve is for $M=420$ MeV, the dashed one for $400$ MeV,
 and the dotted one for $450$ MeV.  The strange quark mass is 
$m_{\rm s}=180$ MeV.  The experimental data are taken from 
SAMPLE~\cite{SAMPLE00s}.  The preferred curve is the one for $M=420$
MeV. 
\newpage

\includegraphics[height=12cm]{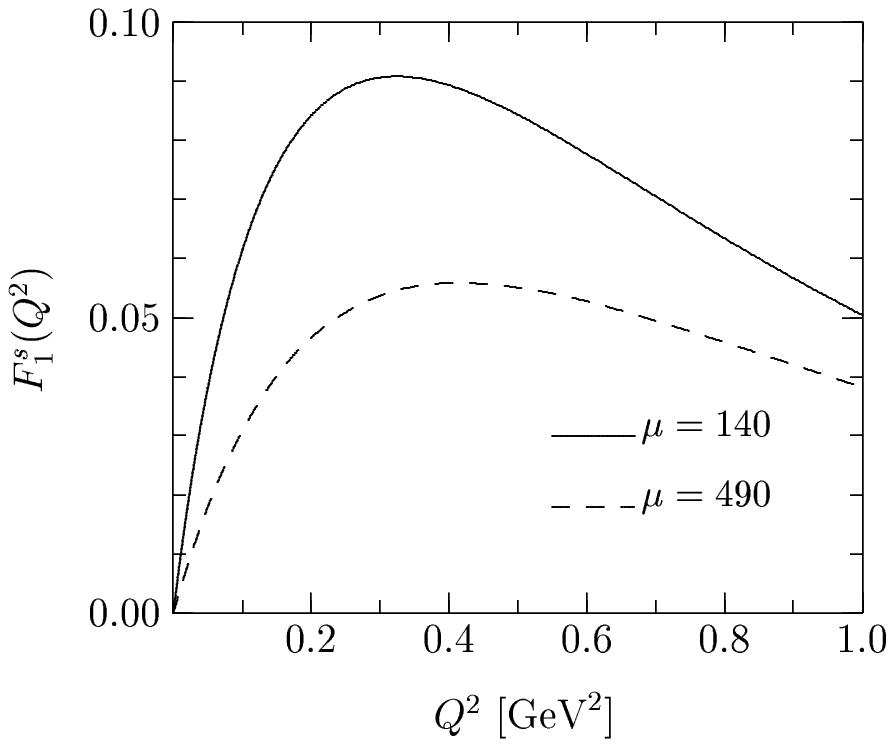}
\vspace{0.8cm}

\noindent {\bf FIG.2}: The form factor $F_1^s$ as a function of $Q^2$.
The solid curve and dashed one represent the results for the kaon
($\mu=490$ MeV) and pion ($\mu=140$ MeV) asymptotic tails, respectively.  The
constituent quark mass is $M=420$ MeV. 
\newpage 
\includegraphics[height=12cm]{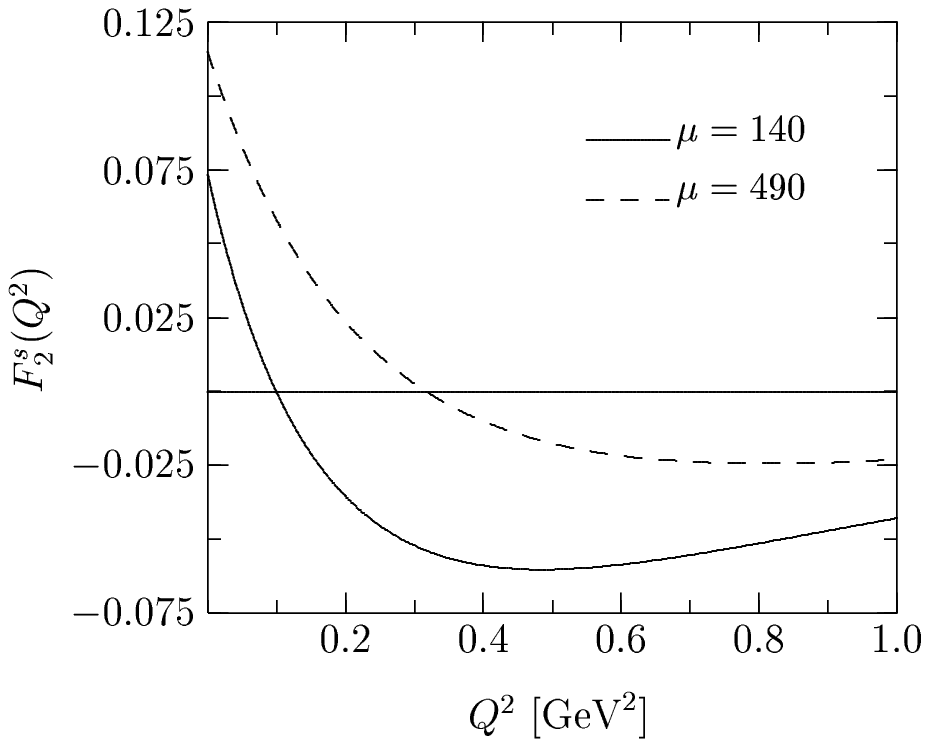}
\vspace{0.8cm}

\noindent {\bf FIG.3}: The form factor $F_2^s$(in physical n.m.) as a
function of $Q^2$.  The solid curve and dashed 
one represent the results for the kaon ($\mu=490$ MeV)  
and pion ($\mu=140$ MeV) asymptotic tails, respectively.  The
constituent quark mass is $M=420$ MeV. 
\newpage 

\centerline{\large \bf TABLES}
\vspace{1.2cm}

\begin{table}[ht]
\caption{Strange form factors: The prediction for the G0 experiment.  
The constituent quark mass $M$ is chosen to be $420$ MeV.  The range 
represents two different results with the pion and kaon tails, 
respectively, indicating the systematic error of the model.}
\begin{tabular}{|c|c|c|c|c|}
\hline
 & \multicolumn{2}{c|}{$\theta=10^\circ$ }
& \multicolumn{2}{c|}{$\theta=108^\circ$}
\\ \cline{1-5}
$Q^2\;[{\rm GeV}^2]$ & $\beta$ & $G_E^{{\rm s}} + \beta G_M^{{\rm
    s}}$ $(\mu = m_{\pi}\sim m_{\rm K})$ & $\beta$ 
&  $G_E^{{\rm s}} + \beta G_M^{{\rm s}}$ 
$(\mu = m_{\pi}\sim m_{\rm K})$
\\ \hline
$0.16$  & $0.13$ & $0.09\sim 0.05$ & $0.63$ & $0.11\sim 0.09$ \\
$0.24$  & $0.20$ & $0.10\sim 0.06$ & $0.99$ & $0.14\sim 0.11$ \\
$0.325$ & $0.26$ & $0.11\sim 0.07$ & $1.31$ & $0.14\sim 0.13$ \\
$0.435$ & $0.35$ & $0.11\sim 0.07$ & $1.81$ & $0.15\sim 0.14$ \\
$0.576$ & $0.47$ & $0.10\sim 0.07$ & $2.49$ & $0.14\sim 0.14$ \\
$0.751$ & $0.61$ & $0.08\sim 0.06$ & $3.35$ & $0.12\sim 0.13$ \\
$0.951$ & $0.81$ & $0.07\sim 0.06$ & $4.62$ & $0.11\sim 0.12$ \\
\hline
\end{tabular}
\end{table}

\begin{table}[ht]
\caption{Singlet form factors: The prediction for the G0 experiment.  
The constituent quark mass $M$ is chosen to be $420$ MeV.  The range 
represents two different results with the pion and kaon tails, 
respectively, indicating the systematic error of the model.}
\begin{tabular}{|c|c|c|c|c|}
\hline
 & \multicolumn{2}{c|}{$\theta=10^\circ$ }
& \multicolumn{2}{c|}{$\theta=108^\circ$}
\\ \cline{1-5}
$Q^2\;[{\rm GeV}^2]$ & $\beta$ & $G_E^{0} + \beta G_M^{0}$ 
$(\mu = m_{\pi}\sim m_{\rm K})$& 
$\beta$ &  $G_E^{0} + \beta G_M^{0}$ 
$(\mu = m_{\pi}\sim m_{\rm K})$
\\ \hline
$0.16$  & $0.13$ & $2.38\sim 2.53$ & $0.63$ & $3.05\sim 3.20$ \\
$0.24$  & $0.20$ & $2.13\sim 2.33$ & $0.99$ & $3.04\sim 3.27$ \\
$0.325$ & $0.26$ & $1.90\sim 2.14$ & $1.31$ & $2.91\sim 3.22$ \\
$0.435$ & $0.35$ & $1.65\sim 1.92$ & $1.81$ & $2.80\sim 3.21$ \\
$0.576$ & $0.47$ & $1.39\sim 1.69$ & $2.49$ & $2.63\sim 3.15$ \\
$0.751$ & $0.61$ & $1.13\sim 1.45$ & $3.35$ & $2.39\sim 3.03$ \\
$0.951$ & $0.81$ & $0.92\sim 1.24$ & $4.62$ & $2.21\sim 2.96$ \\
\hline
\end{tabular}
\end{table}

\begin{table}[ht]
\caption{Strange form factors: The prediction for the A4 experiment.  
The constituent quark mass $M$ is chosen to be $420$ MeV.  The range 
represents two different results with the pion and kaon tails, 
respectively, indicating the systematic error of the model.}
\begin{tabular}{|c|c|c|c|c|}
\hline
 & \multicolumn{2}{c|}{$\theta=35^\circ$ }
& \multicolumn{2}{c|}{$\theta=145^\circ$}
\\ \cline{1-5}
$Q^2\;[{\rm GeV}^2]$ & $\beta$ & $G_E^{{\rm s}} + \beta G_M^{{\rm
    s}}$ $(\mu = m_{\pi}\sim m_{\rm K})$& $\beta$ 
&  $G_E^{{\rm s}} + \beta G_M^{{\rm s}}$
$(\mu = m_{\pi}\sim m_{\rm K})$
\\ \hline
$0.10$  & $0.099$ & $0.07 \sim 0.04$ & $-$ & $-$ \\
$0.227$ & $0.22$ & $0.10\sim 0.06$ & $4.07$ & $0.28\sim 0.32$ \\
$0.47$ & $-$ & $-$ & $8.963$ & $0.33\sim 0.42$ \\
\hline
\end{tabular}
\end{table}

\begin{table}[ht]
\caption{Singlet form factors: The prediction for the A4 experiment.  
 The constituent quark mass $M$ is chosen to be $420$ MeV.  The range 
represents two different results with the pion and kaon tails, 
respectively, indicating the systematic error of the model.}
\begin{tabular}{|c|c|c|c|c|}
\hline
 & \multicolumn{2}{c|}{$\theta=35^\circ$ }
& \multicolumn{2}{c|}{$\theta=145^\circ$}
\\ \cline{1-5}
$Q^2\;[{\rm GeV}^2]$ & $\beta$ & $G_E^{0} + \beta G_M^{0}$ 
$(\mu = m_{\pi}\sim m_{\rm K})$ 
& $\beta$ &  $G_E^{0} + \beta G_M^{0}$
$(\mu = m_{\pi}\sim m_{\rm K})$
\\ \hline
$0.10$  & $0.099$ & $2.61\sim 2.72$ & $-$ & $-$ \\
$0.227$  & $0.22$ & $2.21\sim 2.40$ & $4.07$ & $6.72\sim 7.05$ \\
$0.47$ & $-$ & $-$ & $8.963$ & $7.91 \sim 9.04$ \\
\hline
\end{tabular}
\end{table}

\begin{table}[ht]
\caption{Strange form factors: The prediction for the HAPPEX II experiment.  
The constituent quark mass $M$ is chosen to be $420$ MeV.  The range 
represents two different results with the pion and kaon tails, 
respectively, indicating the systematic error of the model.}
\begin{tabular}{|c|c|c|}
\hline
 & \multicolumn{2}{c|}{$\theta=6^\circ$} 
\\ \cline{1-3}
$Q^2\;[{\rm GeV}^2]$ & $\beta$ & $G_E^{{\rm s}} + 
\beta G_M^{{\rm s}}$ $(\mu = m_{\pi}\sim m_{\rm K})$
\\ \hline
$0.11$  & $0.09$ & $0.07\sim 0.04$ \\
\hline
\end{tabular}
\end{table}
\begin{table}[ht]
\caption{Singlet form factors: The prediction for the HAPPEX II experiment.  
The constituent quark mass $M$ is chosen to be $420$ MeV.  The range 
represents two different results with the pion and kaon tails, 
respectively, indicating the systematic error of the model.}
\begin{tabular}{|c|c|c|}
\hline
 &  \multicolumn{2}{c|}{$\theta=0.09^\circ$} 
\\ \cline{1-3}
$Q^2\;[{\rm GeV}^2]$ & $\beta$ & $G_E^{0} + 
\beta G_M^{0}$ $(\mu = m_{\pi}\sim m_{\rm K})$
\\ \hline
$0.11$  & $0.09$ & $2.55\sim 2.62$ \\
\hline
\end{tabular}
\end{table}

\end{document}